\documentclass{article}
\begin{document}
\newcommand{\QED}{{\hspace*{\fill}\rule{2mm}{2mm}\linebreak}}
\newcommand{\ft}[2]{{\textstyle\frac{#1}{#2}}}
\newcommand{\cmap}{{\bf c}-map}
\def\bigone{{\hbox{1\kern -.23em{\rm l}}}}
\newsavebox{\uuunit}
\sbox{\uuunit}
    {\setlength{\unitlength}{0.825em}
     \begin{picture}(0.6,0.7)
        \thinlines
        \put(0,0){\line(1,0){0.5}}
        \put(0.15,0){\line(0,1){0.7}}
        \put(0.35,0){\line(0,1){0.8}}
       \multiput(0.3,0.8)(-0.04,-0.02){12}{\rule{0.5pt}{0.5pt}}
     \end {picture}}
\newcommand {\unity}{\mathord{\!\usebox{\uuunit}}}
\newcommand{\Ka}{K\"ahler}
\newcommand{\qu}{quaternionic}
\newcommand  {\Rbar} {{\mbox{\rm$\mbox{I}\!\mbox{R}$}}}
\newcommand  {\Hbar} {{\mbox{\rm$\mbox{I}\!\mbox{H}$}}}
\newcommand {\Cbar}
    {\mathord{\setlength{\unitlength}{1em}
     \begin{picture}(0.6,0.7)(-0.1,0)
        \put(-0.1,0){\rm C}
        \thicklines
        \put(0.2,0.05){\line(0,1){0.55}}
     \end {picture}}}
\def\IZ{{\hbox{{\rm Z}\kern-.4em\hbox{\rm Z}}}}
\def\IP{\relax{\rm I\kern-.18em P}}
\def\Im{{\rm Im ~}}
\def\Re{{\rm Re ~}}
\newcommand{\Sp}[1]{\mbox{$Sp\left( #1,\Rbar \right) $}}
\newcommand{\symp}[1]{\mbox{$Sp\left( #1,\Rbar \right) $}}
\newcommand{\sinprod}[2]{\mbox{$\langle #1 , #2 \rangle$}}
\newcommand{\eqn}[1]{Eq.~\ref{#1}}
\newcommand{\matrx}[4]{\left(
  \begin{array}{cc}
    {#1} & {#2} \\
    {#3} & {#4}
  \end{array}\right)}
\newcommand{\bbox}{\lower.2ex\hbox{$\Box$}}

\begin{titlepage}
\begin{flushright} KUL-TF-96/23 \\ hep-th/9611112
\end{flushright}
\vfill
\begin{center}
{\LARGE\bf $N=2$ SUPERGRAVITY AND \\[4mm] SPECIAL GEOMETRY}$^1$   \\
\vskip 27.mm  \large
{\bf   Antoine Van Proeyen}$^2$ \\
\vskip 1cm
{\em Instituut voor theoretische fysica}\\
{\em Universiteit Leuven, B-3001 Leuven, Belgium}
\end{center}
\vfill

\begin{center}
{\bf ABSTRACT}
\end{center}
\begin{quote}
The essential elements in the construction of the couplings of
vector multiplets to supergravity using the conformal approach are
repeated. This approach leads automatically to the basic quantities
on which the symplectic transformations, the basic tools for duality
transformations, are defined. A recent theorem about the existence
of a basis allowing for a prepotential is discussed.
\vfill      \hrule width 5.cm
\vskip 2.mm
{\small
\noindent $^1$ Invited talk at the workshop "Gauge Theories, Applied
Supersymmetry and Quantum Gravity", London Imperial College, july
1996.
\\
\noindent $^2$
Onderzoeksdirecteur, N.F.W.O., Belgium;\newline  E-mail:
Antoine.VanProeyen@fys.kuleuven.ac.be}
\end{quote}
\begin{flushleft}
November 1996
\end{flushleft}
\end{titlepage}

\section{Introduction}
The construction of $N=2$ supergravity and its matter couplings
using superconformal tensor calculus was done in 1979--1984
\cite{N2constr,DWLPSVP,symplN2,dWLVP}. Later, other
constructions~\cite{CdAF}
drew the attention to the freedom of choice of coordinates. The
choice which we made in the first construction, called special
coordinates, was
not inherent to the setup, but rather it was the most natural one to
choose. However, this choice has lead to some confusion by the fact
that coordinates and what are now called the symplectic vectors, were
not clearly distinguished. The symplectic vectors occur most
naturally in the conformal construction. It is therefore no surprise
that symplectic transformations in $N=2$ were first found during the
investigation of the general couplings of vector multiplet
couplings~\cite{symplN2}, nearly immediately after the construction
of special \Ka\ geometry itself~\cite{DWLPSVP}. The latter is by
definition the geometry determined by the kinetic terms of the
scalars in the $N=2,\, d=4$ vector multiplets coupled to
supergravity.
Let me also mention that this approach is now also used to construct
the couplings of supergravity to the vector-tensor multiplet, a
coupling which occurs in the context of string
theories~\cite{vecten}.

The superconformal approach has also its shortcomings. One can not
prove that the obtained action is the most general one. So far, no
procedure has been found to construct the most general couplings of
hypermultiplets. Probably some construction
similar to the harmonic superspace~\cite{harmonic} approach is
required. Furthermore it has been found that formulations for vector
multiplet couplings exist which have not been obtained from superspace
or conformal tensor calculus~\cite{f0art}. In fact, the construction
of the action in these approaches starts from a holomorphic
prepotential. In \cite{f0art}, we constructed vector multiplet
couplings for which there exists no prepotential. On the other hand,
it has been found \cite{Moscowspg,sc2defsg}, in the context of
defining special \Ka\ manifolds in an intrinsic way, that any such
manifold can be described by a prepotential function.
The scalar manifold of any model for which no prepotential exists is
also the scalar manifold of models {\em with} a prepotential, the
relation being made by a duality transformation. However, it is
possible that they are physically different because they
allow different groups to be gauged or different couplings to
hypermultiplets.

With all this, I thought that a review putting the above differences
more explicit would be welcome. At the workshop, for
pedagogical reasons I covered for a large part the rigid special geometry.
This part of the talk can be found in~\cite{Moscowspg}, while a more
extensive review of aspects of the vector multiplet couplings in rigid
supersymmetry has been given in~\cite{trsummer}, whose first two appendices
contain our conventions.

\section{Superconformal tensor calculus}  \label{ss:sctencal}
In this section, I will review the construction and the
ingredients entering in the couplings of vector multiplets to supergravity.
Let it be clear that our aim is not to construct actions
invariant under the superconformal group, but the class of actions
(at most quadratic in derivatives of fields) which are invariant
under the subgroup of the latter: the $N=2$ super-Poincar\'e group,
possibly extended by Yang-Mills gauge invariances.

The principle of superconformal tensor calculus is that one starts
with defining multiplets of the larger group, and then at the very
end breaks it down to the super-Poincar\'e group. This, at the first
sight cumbersome, procedure in fact simplifies life. The
Poincar\'e supersymmetric action contains many terms whose origin is
not clear, unless one recognises how they combine in the
superconformal setup in natural objects (e.g. superconformal
covariant derivatives or auxiliary fields). Also, as we will see,
the symplectic vectors, crucial for the understanding of duality
symmetries, have a natural origin in the superconformal setup.

\subsection{Example without supersymmetry}
The conformal group contains translations $P^a$, Lorentz rotations
$M^{ab}$, Weyl dilatations $D$ and special conformal transformations
$K^a$. These are realised on a 'Weyl multiplet', consisting of the
independent fields $e_\mu^a$, called the vierbein, and $b_\mu$, the
gauge field of dilatations. The gauge fields of the Lorentz rotations
and special conformal transformations are composite fields, resp.
\begin{eqnarray}
\omega_\mu{}^{ab}&=&-2e^{\nu[a}\partial_{[\mu}e_{\nu]}{}^{b]}
+e^{a\rho}e^{b\sigma}
e_{\mu c}\partial_{[\rho}e_{\sigma]}^c +2 e_\mu^{[a}b^{b]}\nonumber\\
f_{\mu a}&=& \ft14 \left(-R_{\mu a}  +\ft16 e_{\mu a}R\right)
\label{compositegf}
\end{eqnarray}
with
\begin{equation}
R_{\mu\nu}{}^{ab}=2\partial_{[\mu} \omega_{\nu]}{}^{ab}+2\omega_{[\mu}
{}^a{}_c\,\omega_{\nu]}{}^{cb}\ ;\quad
R_{\mu a}= R_{\mu\nu ba}e^{\nu b}\ ;\quad R=R_{\mu a} e^{\mu a}\ .
\end{equation}
%\begin{eqnarray}
%R_{\mu\nu}{}^{ab}&=&2\partial_{[\mu} \omega_{\nu]}{}^{ab}+2\omega_{[\mu}
%{}^a{}_c\,\omega_{\nu]}{}^{cb}\nonumber\\
%R_{\mu a}&=& R_{\mu\nu ba}e^{\nu b}\ ;\qquad R=R_{\mu a} e^{\mu a}\ .
%\end{eqnarray}

The envisaged breakdown from superconformal to super-Poincar\'e
symmetry will have to be done by 'compensator fields', i.e. fields
whose value will be fixed in the course of this symmetry breaking. We
illustrate this first with the simplest example. The simplicity has
the disadvantage that one does not see the usefulness of the
procedure, but the reader who wants to be convinced of that, should compare
equations in super-Poincar\'e and superconformal approaches.

Consider one scalar field coupled in a conformal invariant way.
The scalar field is taken to have Weyl weight~1, which means that
it transforms under the Weyl dilatations with parameter $\Lambda_D$
as $\delta \phi= 1.\phi\Lambda_D$. One then shows that the conformal
invariant action is (see e.g. the exercises of lecture~2 in~\cite{Karpacz})
\begin{equation}
{\cal L}=\ft12 e \phi D_a D^a \phi= e\left( -\ft12 \partial_\mu
\phi\cdot \partial^\mu \phi+\ft1{12}R\,\phi^2\right)\ ,\label{lagrconfsc}
\end{equation}
where a covariant
derivative contains connections for $M$, $D$ and $K$-transformations.
The latter leads, via the expression \eqn{compositegf}, to the
$R$-term. The dilatations are gauge fixed by fixing the value of
$\phi$. Taking it e.g. equal to $\sqrt{6}$, we obtain the Einstein
gravity action with the canonical normalisation.
%Note however, that
%to have positive kinetic energy for the graviton, we should have
%started with the negative of \eqn{lagrconfsc}, i.e. the compensating
%scalar should have negative kinetic energy.
$b_\mu$ has
disappeared in \eqn{lagrconfsc}. This is due to the $K$ special
conformal symmetry under which $\delta b_\mu=e_{\mu a}
\Lambda_{K}^a$, while $\phi$ and $e_\mu^a$ are invariant.
In other words $b_\mu$ can be fixed to zero as $K$-gauge choice.
\subsection{$N=2$ conformal tensor calculus}
For $d=4,\,N=2$ the superconformal group is
\begin{equation}
SU(2,2|N=2)\supset SU(2,2)\otimes U(1)\otimes SU(2)\ .\label{scgN2}
\end{equation}
The bosonic subgroup, which I exhibited, contains, apart from the
conformal group $SU(2,2)=SO(4,2)$, also $U(1)$ and $SU(2)$ factors.
The \Ka ian nature of vector multiplet couplings and the \qu\ nature
of hypermultiplet couplings is directly related to the presence of
these two groups, whose gauge fields will become the \Ka\ and \qu\
connections. On the fermionic side there are the two supersymmetries
$Q^i$ and two special supersymmetries $S^i$, which we will also have to
gauge fix. For the gauge fixing of the superfluous gauge symmetries
one makes use of a vector multiplet and a second compensating
multiplet. For the latter three convenient choices exist. We will
use here  a hypermultiplet. The Weyl multiplet has
as independent fields
\begin{equation}
\{ e_\mu^a,\,b_\mu,\, \psi_\mu^i,\,A_\mu,\,{\cal V}_\mu{}^i{}_j,\,
T^-_{ab},\, \chi^i,\, D\}
\end{equation}
These are the gauge fields for general coordinate transformations,
dilatations, $Q^i$, $U(1)$, $SU(2)$ and, as extra fields to close the
algebra, an
antiselfdual antisymmetric tensor, a doublet of fermions and a real
scalar. The gauge field of $S$-supersymmetry is a composite, similar
to those in \eqn{compositegf}, which by the way get extra terms
now~\cite{N2constr,dWLVP}.
As above, $b_\mu$ compensates for the $K$ symmetry. The
vierbein and the gravitinos deliver already all the physical fields
one expects for pure $N=2$ gravity except for the graviphoton, which
sits in the vector compensating multiplet (see below).

A hypermultiplet contains 4 real scalar components and a
doublet of spinors. After choosing the canonical action for this
multiplet, the compensating hypermultiplet completely
disappears as physical multiplet by the gauge choices of $SU(2)$ and
the field equation of $D$, while the field equation of $\chi^i$
eliminates the fermions.\footnote{Although the hypermultiplet disappears
in this way, it can play a non-trivial role. "Gauged $N=2$" and
Fayet-Iliopoulos terms occur when the fields of the hypermultiplets are
taken to transform under a gauge group (with vector multiplets to deliver
the gauge fields). Then the gauge fixing of the scalars of the
compensating hypermultiplet preserves a diagonal subgroup of
the superconformal $SU(2)$ with the
gauge group. As a result also e.g. the gravitino transforms under the
unbroken gauge group, and a Fayet-Iliopoulos term occurs~\cite{symplN2}.
\label{fn:FI}}

For a theory with $n$ physical vector multiplets we thus
introduce $n+1$ vector multiplets to start with.
These consists of fields (with $(w,c)$
weights defining their transformation under the dilatation and $U(1)$
transformations in the superconformal group:
$\delta\phi=w\,\phi\,\Lambda_D +i\,c\,\phi\,\Lambda_{U(1)}$)
\newcommand{\ghe}[3]{\stackrel{\textstyle #1}{\scriptstyle (#2,#3)}}
\begin{equation}
\stackrel {\textstyle(}{\phantom{\scriptstyle (}}
\ghe{X^I_{\phantom{i}}}1{-1},
\ghe{\Omega_i^I}{\frac 32}{-\frac 12},
\ghe{{\cal A}_\mu^I}00, \ghe{Y_{ij}^I}20
\stackrel {\textstyle)}{\phantom{\scriptstyle (}}\qquad
\stackrel{\textstyle\mbox{with}}{\phantom{\scriptstyle (}}\qquad
\stackrel{\textstyle I=0,1,...,n.}{\phantom{\scriptstyle (}}
\label{vectormult}\end{equation}
The action is produced by the chiral superspace integral of a
holomorphic function $F(X)$ as in rigid supersymmetry (see e.g. the
review \cite{trsummer}). However, for invariance under dilatations,
this function $F$ must now be homogeneous of second degree in $X$.
(E.g. $F=(X^1)^3/X^0$ or $F=\sqrt{X^0 (X^1)^3}$ is allowed).

The $Y$-fields (real $SU(2)$ triplets) will be auxiliary in the action.
After gauge choices of dilatations, the $U(1)$ and
$S$-supersymmetry, we will only be left with $n$ complex scalars, $n$
doublets of spinors, and $n+1$ physical vectors, including the
gravitino as mentioned above. The reader can check that all
superfluous symmetries are then broken by gauge choices.

The vectors ${\cal A}_\mu^I$ are gauge fields of an extra $n+1$
dimensional gauge group, for which all the fields of \eqn{vectormult}
are in the adjoint representation, and hypermultiplets may be in
other representations. One of these possibilities was already alluded
to in footnote~\ref{fn:FI}. For the action to be invariant, the
function $F$ is not necessary invariant, because a real quadratic
piece in $F$ does not contribute to the action. Therefore we should
have for the gauge symmetries with parameters $y^I$:~\cite{dWLVP}
\begin{equation}
\delta F\equiv  g F_J f_{IK}{}^J X^K y^I= g y^I C_{I,JK}X^JX^K\ ,
\end{equation}
(for now $F_I, F_{IJ}, ...$ are the derivatives of $F$)
where the $C$ coefficients are real numbers\footnote{For
transformations with non-zero $C$, the \Ka\ potential is not
invariant \cite{Kor}, and this possibility has been omitted in
\cite{Italian}}.

Before taking a gauge choice, typical terms like those in
\eqn{lagrconfsc} occur, now with $R$ multiplied by (after elimination
of the $D$ auxiliary field, see \cite{dWLVP})
$i(\bar X^I F_I - \bar F_I X^I)$. A standard Einstein action can be
obtained by choosing as dilatational gauge fixing
\begin{equation}
i(\bar X^I F_I - \bar F_I X^I)=-X^I N_{IJ}\bar X^J= 1\ ;\qquad
N_{IJ}\equiv 2\,\Im F_{IJ}\label{constraint}
\end{equation}
At this point one could also use the 'string frame' by putting
this equal to an exponential of a dilaton field if the latter is
already well defined. This illustrates the flexibility of the
superconformal approach.
\section{Special \Ka\ geometry}     \label{ss:specialKa}
\subsection{Action, variables and \Ka-Hodge manifolds}
Because of the constraint~\ref{constraint},
the physical scalar fields parametrize an $n$-dimensional
complex hypersurface. It is convenient to write first
\begin{equation}
X^I= a Z^I\ .
\end{equation}
This introduces an extra variable and an extra gauge transformation,
because we can redefine
\begin{equation}
a'= a\, e^{\Lambda_K(Z)}\ ; \qquad Z'^I=Z^I\,e^{-\Lambda_K(Z)} \ .
\label{Kahlertransf}
\end{equation}
For reasons which will become clear soon, we will call this the
'\Ka\ transformation'. Because of the presence of this transformation,
we can choose to have $Z$
invariant under the
dilatation and $U(1)$ transformation, which then act on $a$.
The constraint implies
\begin{equation}
|a|^2=e^K(Z,\bar Z)\equiv i\left( \bar Z^I F_I(Z)-\bar F_I(Z) Z^I\right)
=-\bar Z^I N_{IJ} Z^J\ ,\label{Kahlerpot}
\end{equation}
(due to the homogeneity $F_I(X)=aF_I(Z)$, while $N_{IJ}(X)=N_{IJ}(Z)$),
and the transformation \eqn{Kahlertransf} leads to a possible redefinition
\begin{equation}
K'(Z,\bar Z)= K(Z,\bar Z)+\Lambda_K(Z) + \bar \Lambda_K(\bar Z)\ .
\end{equation}

The action for the scalars is
\begin{equation}
{\cal L}_0=-e\,N_{IJ}{\cal D}_\mu X^I\,{\cal D}^\mu \bar X^J
\qquad\mbox{with}\qquad
{\cal D}_\mu X^I =(\partial_\mu+iA_\mu)X^I\ .
\end{equation}
After elimination of the auxiliary gauge field of the $U(1)$
transformation in the superconformal group,
\begin{equation}
A_\mu=\ft i2 N_{IJ}
\left( X^I\partial_\mu\bar X^J-\partial_\mu X^I\bar X^J\right) \ ,
\label{Amusoln}
\end{equation}
and writing ${\cal L}_0$ in terms of $Z^I$, the phase of $a$
disappears because of the $U(1)$. In other words we can take the
gauge choice $a=\bar a$, which leaves a combination of the $U(1)$ and
the \Ka\ transformation \eqn{Kahlertransf}:
\begin{equation}
2i \Lambda_{U(1)}= \Lambda_K(Z) -\bar \Lambda_K(\bar Z)\ .
\label{decompU1K}
\end{equation}
The action then becomes of the \Ka ian type
\begin{equation}
{\cal L}_0= - \partial_\mu Z^I \, \partial_\mu \bar Z^J\, \frac{\partial}
{\partial Z^I}\frac{\partial}{\partial\bar Z^J}K(Z,\bar Z)\ .
\label{L0Kahler}
\end{equation}
The remaining \Ka\ transformation can e.g. be used to choose one of
the $Z^I$, say $Z^0$, equal to~1. In any case, one can choose the
parametrization of the $n$ physical scalars $z^\alpha$ (with
$\alpha=1,\ldots ,n$) at random, as stressed in
\cite{CdAF}. To preserve the complex structure, already apparent in
\eqn{L0Kahler}, the relations $Z^I(z^\alpha)$ should be holomorphic
functions.  E.g.
a convenient choice for the inhomogeneous coordinates $z^\alpha $ at
$Z^0\neq 0$ are the {\it special} coordinates, defined by $Z^0=1$,
$Z^\alpha = z^\alpha $. These were always used in the articles
before~\cite{CdAF}.
The resulting geometry is known as {\it special} \Ka\ geometry
\cite{DWLPSVP,symplN2,special}. There is one global aspect which is
important, and as far as I know it is the only instant where the
fermion sector comes in. As implied in \eqn{vectormult}, the fermions
transform under the superconformal $U(1)$ factor, and hence, by
\eqn{decompU1K}, under the (finite) \Ka\ transformations:
\begin{equation}
%\delta\Omega_i=-\ft14\Omega_i\,
%\left( \Lambda_K(Z) -\bar \Lambda_K(\bar Z)\right) \ ,\qquad
%\mbox{or finite transformation:}\quad
\Omega_i\rightarrow
e^{-\ft14\left( \Lambda_K(Z) -\bar \Lambda_K(\bar Z)\right)}\Omega_i
\ .
\end{equation}
Then one argues in the same way as for the magnetic monopole
(as nicely explained in \cite{BaggerWittenN1}): the fermion should
remain well defined when going e.g. around a sphere. Therefore in
going around in different patches, the reparametrisations should be
such that $\Lambda_K(Z) -\bar \Lambda_K(\bar Z)$ is $2\pi\, i$ times
an integer. The properly normalised gauge field is then
$A_\mu/(4\pi)$ (using \eqn{decompU1K}), and with \eqn{Amusoln} the
field strength is
\begin{equation}
\Omega_{\mu\nu}=-\frac i{4\pi} \frac{\partial^2 K}{\partial Z^I\partial Z^J}
\partial_{[\mu}Z^I\,\partial_{\nu]}\bar Z^J\ ,
\end{equation}
hence, proportional to the \Ka\ 2-form. \Ka\ manifolds on which the
transitions between coordinates in different patches can be done with
such transformations with integer parameters are called '\Ka -Hodge'.

%On the global manifold there can be different patches where other
%coordinates are used. On overlaps of three regions, the fermion
%should be well defined. This implies, as nicely explained in
%\cite{BaggerWittenN1}, that if $\Lambda_{AB}(Z)$ denotes the \Ka\
%transformation that is necessary to go from region A to B, one
%should have that $i\left( \Lambda_{AB}
%+\Lambda_{BC}+\Lambda_{CA}\right) $, which is a real constant out of
%consistency,  should be $4\pi$ times an integer. This is only
%possible on so-called 'Hodge' manifolds.

Let us still note that the conditions for positive kinetic energies
are that $g_{I\bar J}=\frac{\partial}{\partial Z^I}
\frac{\partial}{\partial \bar Z^J} K$ is positive (with one zero
mode), and that $Z^IN_{IJ}\bar Z^J $ is negative. That determines the
positivity domain of the variables.
\subsection{Duality transformations, intrinsic definition and existence
of the prepotential}
So far, we started from a particular function $F(X)$, but duality
symmetries imply that different functions can lead to the same
manifold. To understand the duality transformations, one has to
consider the coupling to the vectors. This is governed by the complex
symmetric tensor
\begin{equation}
{\cal N}_{IJ} =\bar F_{IJ} +i {{(N_{IN})( N_{JK})Z^N
Z^K}\over{(N_{LM})\ Z^L Z^M}}\ .
\label{Ndef}
\end{equation}
describing the coupling of the vector field strengths\footnote{Its imaginary
part is negative definite in the positivity domain~\cite{BEC}.}.
After some
analysis of the fields equations, it turns out that duality
transformations, leading to equivalent field equations (for abelian
gauge fields), can be obtained by transformations in the group
$Sp(2(n+1), \Rbar)$, and taking the quantisation conditions of
charges into account, these transformations are restricted to integers.
Under such transformations with a matrix
\begin{equation}
{\cal S}=\matrx A B C D \in Sp(2(n+1), \Rbar)\ ,
\end{equation}
${\cal N}$ should change to $(C+D{\cal N})(A+B{\cal N})^{-1}$. This
can be obtained by considering
\begin{equation}
v=\pmatrix {Z^I\cr F_I(Z)}
\end{equation}
as a symplectic vector: $v'= {\cal S}v$.  The \Ka\ potential
\eqn{Kahlerpot} is obviously an invariant. The lower components of
$v'$ are again the derivatives of a scalar function $F'(Z')$ if the
relation $Z'^I(Z)=A^I{}_JZ^J+B^{IJ}F_J(Z)$ is invertible, or in other
words $A^I{}_J +B^{IK}F_{KJ}$ is invertible. $A+B{\cal N}$ is
invertible in the positivity domain ($\Im
{\cal N}<0$). In rigid supersymmetry ${\cal N}_{IJ}=\bar F_{IJ}$, and
$F'$ thus always exists. However, in the supergravity case important
exceptions are known~\cite{f0art}.

This implies that there are formulations where the prepotential $F$
does not exist. These have thus been constructed by starting with a
prepotential $F$, and then making a symplectic transformation.
This is one of the arguments to formulate a
definition of a special \Ka\ manifold
without making reference to such a prepotential. That was already
done by Strominger~\cite{special}, but it turned out
that his definition was not restrictive enough for allowing $N=2$
supergravity~\cite{Moscowspg,sc2defsg}, which should be the decisive
criterion to decide whether
a manifold is special \Ka. A new definition has been given, and it
was proven that all manifolds allowed by this new definition can be
obtained starting from the above construction with a
prepotential~\cite{Moscowspg,sc2defsg}.

One definition of a special \Ka\ manifold is as follows. It is an
$n$-dimensional Hodge-\Ka\
manifold ${\cal M}$ with a positive definite metric and
with ${\cal L}$ a complex line bundle whose first
Chern class equals the \Ka\ form ${\cal K}$. There is a
$\symp{2(n+1)}$ vector bundle ${\cal H}$ over ${\cal M}$, and a
holomorphic section $v(z)$ of ${\cal L}\otimes{\cal H}$, such that the
\Ka\ potential is given by \eqn{Kahlerpot}, and such that
\begin{equation}
\sinprod{v}{\partial_{\alpha} v} = 0\ ;\qquad
\sinprod{\partial_\alpha v}{\partial_{\beta} v} = 0\ . \label{cond3local}
\end{equation}

If the metric is positive definite, one can show \cite{sc2defsg} that
$(\bar X^I,\partial_\alpha X^I)$ is an invertible $(n+1)\times(n+1)$
matrix, and one defines
\begin{equation}
\bar {\cal N}= \left[\bar X^I,\partial_\alpha X^I\right]
\left[(\bar X^I,\partial_\alpha X^I)\right]^{-1}\ ,
\end{equation}
from which one proves ${\cal N}={\cal N}^T$ and $\Im {\cal
N}<0$. If a prepotential exists then this leads to \eqn{Ndef}.

It turns out that the following 4 conditions are equivalent: 1)
$( X^I,\partial_\alpha X^I)$ invertible; 2)
special coordinates are possible; 3)
$e_\alpha^A=\partial_\alpha \left(\frac{X^A}{X^0} \right) $
(for a choice of the coordinate "0", and $A=1,...,n$); 4) a
prepotential $F(X)$ exists.

For the last statement, one first notices that $(X^I/X^0)$ is
independent of $\bar z$, such that from the functions
$F_I(z,\bar z)$, one can define holomorphic functions
\begin{equation}
F_I(X)\equiv X^0 \frac{F_I}{X^0}
\left( z\left(\frac{X^A}{X^0} \right) \right) \ .
\end{equation}
The constraints
\eqn{cond3local}  then
imply
\begin{equation}
\pmatrix{ X^I\cr \partial_\alpha X^I}\partial_{[I}F_{J]} \
\pmatrix{ X^J&\partial_\alpha X^J}=0\ ,
\end{equation}
from which it follows that in
any patch $F_J=\frac{\partial}{\partial X^J}F(X)$ for some $F(X)$.

Finally, we have proven in \cite{sc2defsg} that for any special \Ka\
manifold there exists a symplectic transformation to a symplectic
basis such that the conditions 1) are satisfied.

This implies that all the \Ka\ manifolds which are special, can be
constructed from prepotentials. Indeed for symplectic vectors which
satisfy the necessary conditions, we in general still need a
symplectic transformation to obtain a formulation with a
prepotential, but the \Ka\ potential, defined in \eqn{Kahlerpot}, is
obviously invariant under such a transformation.

On the other hand, such a symplectic transformation is not necessary
an invariance of the action, so the models without prepotential are
necessary for a complete description of the matter couplings of $N=2$
vector multiplets, as it was shown in~\cite{f0art}.

Finally, let me remark that the symplectic transformations as used
above act on the vector define new vectors $\tilde Z^I$. Therefore
one can often associate to these also a coordinate transformation if
the coordinates $z^\alpha$ are defined in terms of $Z^I$ (e.g. the
special coordinates). In the latter case, the symplectic
transformations induce coordinate transformations. As in the early
papers of special geometry one always used special coordinates, the
symplectic transformations were usually considered in combination
with these coordinate transformations. If that is done, then the \Ka\
potential is not necessary invariant. The combined transformations
which leave the \Ka\ potential invariant then define 'duality
symmetries' \cite{symplN2,BEC}. But one should clearly distinguish
these transformations. In \cite{christoine} this has been discussed
in a more general context.

\section*{Acknowledgments}
Part of this work was done in various collaboration. I wish to thank
especially B. de Wit, W. Troost, B. Craps, Fr. Roose, S. Ferrara, P.
Fr\`e, R. D'Auria, A. Ceresole. This work was supported by
the European Commission TMR programme ERBFMRX-CT96-0045.

\newcommand{\Journal}[4]{{#1} {\bf #2}, #3 (#4)}
% Some useful journal names
\newcommand{\NCA}{\em Nuovo Cimento}
\newcommand{\NIM}{\em Nucl. Instrum. Methods}
\newcommand{\NIMA}{{\em Nucl. Instrum. Methods} A}
\newcommand{\NPB}{{\em Nucl. Phys.} B}
\newcommand{\PLB}{{\em Phys. Lett.}  B}
\newcommand{\PRL}{\em Phys. Rev. Lett.}
\newcommand{\PRD}{{\em Phys. Rev.} D}
\newcommand{\ZPC}{{\em Z. Phys.} C}
\def\IJMPA{{\em Int. J. Mod. Phys.}A}
\def\CMP{\em Comm. Math. Phys.}
\def\CQG{\em Class. Quantum Grav.}

\end{document}